\documentclass[12pt,preprint]{aastex}

\slugcomment{Accepted for publication in the ApJ}

\shorttitle{Vela - D Knot}
\shortauthors{Sankrit, Blair and Raymond}

\begin{document}

\title{Optical and Far-UV Spectroscopy of Knot D in 
the Vela Supernova Remnant}

\author{Ravi Sankrit,  William P. Blair}
\affil{The Johns Hopkins University}
\affil{Department of Physics and Astronomy,
3400 N. Charles St.,  Baltimore, MD 21218}
\email{ravi@pha.jhu.edu, wpb@pha.jhu.edu}
\and
\author{John C. Raymond}
\affil{Harvard-Smithsonian Center for Astrophysics,
       60, Garden St., Cambridge, MA 02138}
\email{jraymond@cfa.harvard.edu}


\begin{abstract}

We present spectra of optical filaments associated with the X-ray knot
D in the Vela supernova remnant.  It has been suggested that Knot D is
formed by a bullet of supernova ejecta, that it is a break-out of the
shock front of the Vela SNR, and also that it is an outflow from
the recently discovered remnant RXJ0852.0-4622.  We
find that Knot D is a bow shock propagating into an interstellar cloud
with normal abundances and typical cloud densities (n$_{\rm{H}} \sim$ 4
-- 11~cm$^{-3}$).  Optical longslit spectra show that the
[S~II]~$\lambda\lambda$6716,6731 to H$\alpha$ line ratio is greater
than unity, proving that the optical filaments are shock excited.  The
analysis of far-ultraviolet spectra obtained with the Hopkins
Ultraviolet Telescope and with the Far Ultraviolet Spectroscopic
Explorer (FUSE) LWRS aperture show that slower shocks
($\sim100$~km~s$^{-1}$) produce most of the low ionization lines such
as O~III]~$\lambda$1662, while faster shocks ($\sim180$~km~s$^{-1}$)
produce the O~VI~$\lambda\lambda$1032,1038 and other high ionization
lines.  C~III and O~VI lines are also detected in the FUSE MDRS
aperture, which was located on an X-ray bright region away from the
optical filaments.  The lines have two velocity components consistent
with $\sim150$~km~s$^{-1}$ shocks on the near and far sides of the
knot.  The driving pressure in the X-ray knot, $P/k_{\rm{B}} \sim
1.8\times10^7$~cm$^{-3}$~K, is derived from the shock properties.  This
is over an order of magnitude larger than the characteristic X-ray
pressure in the Vela SNR.  The velocity distribution of the emission
and the overpressure support the idea that Knot D is a bow shock around
a bullet or cloud that originated near the center of the Vela remnant.

\end{abstract}


\section{Introduction}

In an X-ray image obtained with ROSAT, the Vela Supernova Remnant (SNR)
is roughly circular with a diameter $\sim8\arcdeg$ \citep{asc95}.  At a
distance of 250~pc \citep{cha99} this corresponds to about 35~pc.  Six
bow shaped ``knots'' of emission lie beyond the nominal circumference
of the remnant.  The shape and location of these knots led Aschenbach
et al.~to suggest that they were due to ejecta ``bullets'' that had
overtaken the blast wave.  Of the six features, the closest to the
remnant and also the brightest in X-ray is ``Knot D''\@.  The optical
nebula RCW~37 \citep{rod60}, lies along the outer edge of Knot D\@.
This association with bright optical filaments also distinguishes Knot
D from the other five knots.

Although it is grouped with the ejecta bullets, there are two
other explanations for the origin of Knot D\@.  One scenario,
suggested by \citet{plu02} is that Knot D is a shock that has broken
out from the Vela SNR due to inhomogeneities in the ambient medium.
They analyzed \textit{Chandra}-ACIS spectra of Knot D and found that
the oxygen abundance was solar and that the neon abundance was modestly
enhanced, and they found no evidence for abundance differences at
different locations in the knot.  These results as well as the
morphology of the X-ray emission led \citet{plu02} to favor a shock
break-out origin for Knot D\@.  The other suggestion \citep{red00,
red02} is that Knot D is associated with RXJ0852.0-4622, a remnant that
in projection is within the boundaries of the Vela SNR\@.  They
analyzed optical echelle data of RCW~37 and found a velocity split near
the bright optical edge in both [S~II] and [O~III] emission, and an
almost complete velocity ellipse across the nebula in the latter.  They
inferred that the geometry of the optical nebula is either an
incomplete funnel or a wavy sheet.  Based on the kinematic structure of
emission as well as the optical and X-ray morphologies of the nebula
they suggested that an outflow from RXJ0852.0-4622 impacted the
pre-existing wall of the Vela SNR to produce Knot D\@.  As they point
out, for their model to work both remnants have to be at about the
same distance \citep{red02}.

Regardless of its origin, it is known that Knot D is a source of shock
excited emission.  \citet{bla95} presented far-ultraviolet observations
of the region obtained with the \textit{Voyager 2} Ultraviolet
Spectrometer.  Strong emission from C~III~$\lambda$977 and
O~VI~$\lambda\lambda$1032,1038 was detected in the spectrum.  The
authors noted the sheet-like morphology of the filaments and the
spatial coincidence of the X-ray and optical emission and argued that
these were characteristics of a shock-cloud interaction (such as had
been seen in the Cygnus Loop SNR) and that this interaction produced
the ultraviolet lines as well as the optical emission.  They suggested
that the large \textit{Voyager} FOV included shocks with a range of
velocities, and that $\sim120$~km~s$^{-1}$ shocks were responsible for
the C~III emission and 160--300~km~s$^{-1}$ shocks produced the O~VI
emission.

In this paper we take a detailed look at the shock-cloud interaction
and determine the shock wave paramaters.  Our results throw some light
on the nature of Knot D, but do not unambiguously determine its
origin.  We use ultraviolet spectra obtained with the \textit{Hopkins
Ultraviolet Telescope} (HUT) and with the \textit{Far Ultraviolet
Spectroscopic Explorer} (FUSE).  These data have much higher spectral
resolution than the \textit{Voyager} spectrum and the aperture FOVs are
over a thousand times smaller.  We also present optical longslit
spectra of the filaments.  We confirm that the optical emission is
shock excited.  By comparing the measured far-ultraviolet emission line
strengths with shock model predictions, we find the properties of the
shocks around Knot D\@.  The kinematic distribution of the emission
around the bow shock is clearly revealed in the high resolution FUSE
data.  The observations are described in \S2, and the results presented
in \S3.  Then, in \S4 we analyze the results and present our
interpretation.  The last section, \S5 summarizes the central results
of this work and presents some of the issues yet to be answered.

\section{Observations}

Images of the region of interest are shown in Figure \ref{f_image}.
The top panel is a three color image of the southern half of Knot D\@.
Narrowband H$\alpha$ and [O~III]~$\lambda$5007 images (first presented
by \citet{bla95}) are shown in red and green, respectively.  The blue
is the \textit{Chandra} X-ray image, presented by \citet{plu02}
(and kindly provided to us by the lead author).  Overlaid on the image are
the locations of the FUSE LWRS and MDRS apertures.  The bottom panel is
a two color image of a region around the bright filament.  Only the
optical emission is shown (H$\alpha$ in red, [O~III] in green)
overlaid with the HUT and FUSE LWRS aperture positions.

The HUT observations were made during the Astro-2 space shuttle mission
on 1995 March 15.  (The basic HUT design is described by \citet{dav92},
and the improvements to the instrument for Astro-2 by \citet{kru95}.)
The wavelength coverage of HUT was 820 -- 1840\AA\ with a resolution of
$\sim$~3\AA\@.  The 56\arcsec$\times$10\arcsec\ aperture was centered
at $\alpha_{2000}=09^{\rm{h}} 00^{\rm{m}} 24\fs30$,
$\delta_{2000}=-45\arcdeg\ 54\arcmin\ 31\farcs5$, and placed at a
position angle of 18\arcdeg\, which aligned the aperture with the
optical filament.  After approximately 1000 s of exposure time at this
position, the aperture was offset to a location $\sim$30\arcsec\ East,
perpendicular to the aperture's long dimension.  We refer to these
positions as P1 and P2, respectively (Figure \ref{f_image}b).  The data
were obtained in a time-tag mode which allowed the separation of
orbital day and night photons.  Since the daytime photons are
contaminated by airglow emission, only the nighttime data are used
here.  The effective exposure times are 932~s for P1 and 1162~s for
P2.

The HUT data were processed using an IRAF\footnote{IRAF is distributed
by the National Optical Astronomy Observatories, which are operated by
the Association of Universities for Research in Astronomy, Inc., under
cooperative agreement with the National Science Foundation.} package
originally developed to process HUT data from the Astro-1 mission, and
described by \citet{kru99}.  The flux calibration applied to these data
is based on in-orbit observations of white dwarf stars fitted with
theoretical models.  Other corrections, such as pulse persistence, dark
count and Ly$\alpha$ scattering have been characterized and included
here \citep{kru95,kru99}.

The FUSE observation (Program ID B1080201) was obtained on 2001 April 2
as part of a Guest Investigator project to study the Vela-Puppis
region.  Eleven exposures with a total integration time of 14,133 s
were obtained with the low resolution (LWRS)
30\arcsec$\times$30\arcsec\ aperture centered at
$\alpha_{2000}=09^{\rm{h}} 00^{\rm{m}} 24\fs30$,
$\delta_{2000}=-45\arcdeg\ 54\arcmin\ 31\farcs5$.  This position is the
same as the center of the HUT P1 observation, and lies on the bright
optical filament (Figure~\ref{f_image}b).  Data were obtained
simultaneously through the medium resolution (MDRS)
20\arcsec$\times$4\arcsec\ aperture.  The MDRS aperture was located
208\arcsec\ away, 24\arcdeg\ West of North from the LWRS aperture on an
X-ray bright region away from the bright optical emission
(Figure~\ref{f_image}a).

The wavelength range covered by FUSE is 905 -- 1187\AA.  In this
paper we present data from segments SiC2A ($\sim$905 -- 1000\AA), LiF1A
($\sim$1000 -- 1100\AA) and LiF2A ($\sim$1100 -- 1187\AA), the segments
with largest effective area in their wavelength ranges \citep{moo00}.
The data in other segments were used for comparison to distinguish
between real features and possible detector artifacts.  The raw data
from all the exposures were combined and the pipeline CalFUSE version
1.8.7 was used to produce calibrated spectra.  A shift was applied to
the LiF1A and SiC2A flux vectors to line up the geocoronal emission
(Ly$\beta$ and Ly$\gamma$ respectively for the two channels) at the
appropriate wavelength in the heliocentric frame of reference.  This
was done for both LWRS and MDRS spectra.  The source is an extended
emission object and fills the apertures.  For such a target, the
effective spectral resolution is a combination of the instrumental
resolution ($\sim$0.05\AA), the slit width, and a further degradation
caused by detector astigmatism \citep{sah00}.  The width of the airglow
lines in a spectrum is a measure of the effective spectral resolution.
In our data, the effective spectral resolutions are $\sim$0.34\AA\ for
the LWRS data and $\sim$0.09\AA\ for the MDRS data.  (These numbers
correspond to about 100~km~s$^{-1}$ and 30~km~s$^{-1}$, respectively,
at 1000\AA\@.)

The longslit optical spectra presented here were obtained in 1989
February at Las Campanas Observatory.  The observations used the du
Pont 2.5~m telescope, the Modular Spectrograph and an $800\times800$
pixel TI CCD\@.  A 600~line~mm$^{-1}$ 5000\AA\ blaze grating was used
with an 85~mm camera, providing spectral coverage, 4750 -- 7150\AA\ at
3.0\AA\ per pixel.  The slit width was 2\arcsec\ and the effective
spectral resolution was 8\AA\@.  The spatial scale along the slit was
0\farcs8~pixel$^{-1}$, with a usable slit length of 7\arcmin.  The slit
was placed along the East-West direction, passing through the optical
filaments $\sim1$\arcsec\ below the slit center of the HUT position P1.

The data were reduced using standard IRAF procedures, including bias
subtraction, flat fielding, image rectification and extraction of one
dimensional spectra.  The background was determined from a portion of
the slit lying outside the remnant and subtracted from the spectra.
The flux calibration was done using observations of standard stars from
the list of \citet{sto83}.  Fiducial stars along the slit, and the
known spatial scale were used to extract regions of the long slit data
corresponding to the intersection of the slit with the HUT apertures.
The optical images (Figure \ref{f_image}b) show that the emission is
fairly uniform within the apertures, so the optical spectra should be
representative of the overall emission within the HUT apertures at P1
and P2.

\section{Discussion}

\subsection{Optical Imagery and Spectra}

The optical narrowband images show a striated emission structure.
There are bands of green (strong [O~III]), yellow (strong [O~III] and
H$\alpha$) and red (strong H$\alpha$) running parallel to the shock
front (Figure~\ref{f_image}b), indicating a systematic variation of the
H$\alpha$ to [O~III] flux ratio.  The H$\alpha$ emission is at its
brightest in the region between about 50\arcsec\ and 70\arcsec\ behind
the leading [O~III] edge.  The [O~III] emission reaches its peak at
about the same location as the H$\alpha$ ($\sim50\arcsec$ behind the
edge).  In the green ([O~III] dominated) band of emission, the [O~III]
flux is about 40\% of its peak value, while the H$\alpha$ flux remains
below 20\% of its peak value.

The brightest optical filaments lie approximately parallel to the
Southeast edge of the X-ray emitting region (Figure~\ref{f_image}a) and
there is a gap of about 1\arcmin\ between the X-ray edge and the
filaments.  Observations of other shock-cloud interactions in the
Cygnus Loop \citep{hes86, dan00} and in Vela \citep{ray97, san01} have
shown that optical filaments are typically coincident with the edges of
bright X-ray features, and not separated as in Knot D\@.  So though the
association of X-ray and optical regions is expected in SNR shocks, the
existence of a gap between the two is surprising.

The optical spectra of the regions of bright and faint H$\alpha$ (P1
and P2, respectively) are shown in Figure~\ref{f_opt}.  We measured the
integrated line fluxes using the IRAF task, \texttt{splot}, and
calculated the surface brightnesses.  We derive the color excess from
the observed ratio between between the H$\alpha$ and H$\beta$ fluxes.
The intrinsic ratio between the two lines for shock excited emission is
3.0 \citep{ray79}.  The observed ratio at P1,
I$_{\rm{H}\alpha}$/I$_{\rm{H}\beta}\simeq3.4$, implies that $E_{B-V} =
0.1$, a value consistent with those obtained for other regions in Vela
\citep{wal90}.  At position P2, we only have an upper limit for the
H$\beta$ flux and cannot estimate the color excess.  Therefore we
assume that the P1 value applies to the whole region.  We used the
extinction curve suggested by \citet{fit99}, and total to visual
selective extinction $R = 3.1$ to calculate correction factors for all
the lines.  The surface brightnesses of the optical lines, corrected
for interstellar extinction are presented in Table~\ref{tbl_opt}.

The brightness ratio, I$_{\rm{[S~II]}}$/I$_{\rm{H}\alpha}$ is greater
than 1 at both positions P1 and P2.  The high ratios imply that the gas
is collisionally excited and confirm that the optical filaments are due
to shocks.  As expected from the narrowband images, the optical spectra
at the two positions differ significantly in the strength of [O~III]
emission relative to emission from the lower ionization lines.  The
[O~III]~$\lambda\lambda$4959,5007 flux is about 3 times weaker at
position P2 than at P1, whereas all the other lines are about an order
of magnitude weaker at P2.  Some lines, including
[O~I]~$\lambda\lambda$6300,6364 are not detected at the latter
position.  This difference suggests that at P2, the shock is
``incomplete'' - it has not yet swept up enough material for the
post-shock gas to have recombined and produced strong H$\alpha$,
[S~II] and other low ionization lines.

\subsection{HUT Spectra}

The HUT spectra of the two positions in Knot D contain numerous
emission lines expected from SNR shocks (Figure~\ref{f_hut}).  As
expected from the \textit{Voyager} UVS spectrum \citep{bla95},
C~III~$\lambda$977 and O~VI~$\lambda\lambda$1032,1038 are the strongest
lines below 1200\AA\@.  Other lines such as N~III~$\lambda$991, which
were lost in the low resolution \textit{Voyager} spectrum, are detected
in these HUT data.  Longward of 1200\AA, the strongest lines are
C~IV~$\lambda$1549, O~III]~$\lambda$1663 and N~III]~$\lambda$1751.
Many (but not all) of the detected lines are marked in
Figure~\ref{f_hut}.  One surprise is that N~V~$\lambda$1240 is
relatively weak: in spectra of radiative shocks in the Cygnus Loop the
line is comparable in strength to the C~IV line \citep{bla91}.

The integrated line fluxes in the spectra were measured using the IRAF
tasks \texttt{splot} and the more sophisticated \texttt{specfit}
\citep{kri94}, the latter for blended lines.  The pair of lines,
Si~IV~$\lambda$1402 and O~IV]~$\lambda$1403 could not be deblended
automatically.  Therefore, it was assumed that the Si~IV~$\lambda$1402
flux was exactly half the Si~IV~$\lambda$1394 flux (valid for optically
thin emission).  This amount was subtracted from the measured flux of
the 1403\AA\ feature and added to the 1394\AA\ line.  The surface
brightnesses of the lines, corrected for interstellar extinction, are
listed in  Table~\ref{tbl_hut} along with the correction factors.  It
should be noted that the interstellar extinction curve at wavelengths
below 1000\AA\ is poorly known and is the main source of uncertainties
in the intrinsic brightness of the shortest wavelength lines.

The line strengths at P1 are about twice those at P2, but apart from
that overall factor the spectra at the two positions are similar.  The
dramatic differences in the relative intensities of optical lines are
not present for the ultraviolet lines.  This implies that the shock
conditions are similar at the two positions.  For instance
I$_{\rm{O~VI}1032,1038}$/I$_{\rm{O~III]}1664}$, a ratio that is very
sensitive to the maximum temperature reached behind the shock, is $\sim
1.3$ at P1 and $\sim 1.2$ at P2.  Thus the far-ultraviolet spectra
support the idea that the difference in the optical emission between P1
and P2 is due to different levels of shock completeness, rather than
differences in shock velocity.  The main difference between the HUT P1
and P2 spectra is the ratio between the two O~VI doublet lines:
I$_{1038}$/I$_{1032}\simeq0.76$ at P1 and $\simeq0.61$ at P2.  The
ratio approaches 0.5 when the line optical depth tends to zero, and
approaches 1.0 when the line is optically thick.  The measured ratios
show that the line fluxes are affected by resonance line scattering and
that the effect is stronger at P1 than at P2.

The ratio between the O~VI line fluxes can be affected by absorption of
the 1038\AA\ line by molecular hydrogen (the Lyman band transitions,
R(1)$_{5-0}$ and P(1)$_{5-0}$, at 1037.15\AA\ and 1038.16\AA,
respectively) and by C~II* (1037.02\AA).  The ratio can also be
affected by self-absorption of the lines by O~VI in the ISM.
\citet{jen76} find the column of low velocity H$_2$ in the J=1 state
$N_{J=1} \simeq 10^{14.5}$, 10$^{15.5}$, and 10$^{17.8}$ towards three
stars behind Vela.  Even the highest of these would absorb only about
9\% of the flux from a 150~km~s$^{-1}$ width line centered at zero velocity.
The more typical lower values for the column would have negligible
effect on the O~VI~$\lambda$1038 flux.  The C~II* line is stronger, but
is at $-173$~km~s$^{-1}$ relative to the O~VI~$\lambda$1038 rest
wavelength.  The velocity profile of the O~VI lines in the FUSE LWRS
spectrum (see below) shows that C~II* absorption would not reduce the
O~VI flux significantly.  The O~VI column out to the distance of Vela
(250~pc) is $\sim10^{13}$~cm$^{-2}$ \citep{jen01}, which would have
negligible affect on the line fluxes.  Also, Knot D lies beyond the
remnant boundary, and it is unlikely that other shocked clouds
associated with Vela lie in front of it.  Therefore, the ratios
measured in the HUT spectra should not be modified significantly 
by foreground absorption.

\subsection{FUSE Spectra}

\subsubsection{LWRS Spectrum}

The FUSE LWRS spectrum is shown in Figure~\ref{f_fuse}.  The strongest
SNR lines in the FUSE bandpass are C~III~$\lambda$977 and
O~VI~$\lambda\lambda$1032,1038.  These lines are allowed to go
off-scale in the plot in order to show a number of the weaker lines
present in the spectrum.  The weak lines are well separated from each
other and are easily identified.  Furthermore, SNR lines can be
distinguished from airglow lines by comparing the total spectrum with a
spectrum screened to include only orbital night data.  Many of these
weaker lines have been detected in SNR spectra for the first time by
FUSE\@.  For example, this paper reports the first such detection of
S~III~$\lambda$1077.  Note that the region around
He~II$+$[N~II]~$\lambda$1085 is not included in the plots.  This range
lies at the edge of only one detector segment \citep{sah00}, and the
telescope effective area is very small in this region.

The total line fluxes in the LWRS spectrum were obtained by trapezoidal
integration of the flux vector with suitable background subtraction.
The strong lines contain several thousand counts, so the random errors
in their measured fluxes are low ($\sim$~1\%).  The errors are about
6\% for some of the weaker lines, such as S~III~$\lambda$1015 and
S~IV~$\lambda$1063.  For the weakest line fluxes, particularly those
detected in the SiC~2A segment (e.g.~S~VI~$\lambda\lambda$933, 944),
uncertainties in the background placement contribute to the total
error, which is about 20\%.  The uncertainty in the absolute flux
calibration is about 10\% \citep{sah00}, which dominates the
statistical errors in the measured fluxes of the strong lines and is
comparable to those of the weak lines.  The surface brightnesses of the
lines, corrected for interstellar extinction, are presented in
Table~\ref{tbl_lwrs}.

The fluxes of the strong lines (C~III and O~VI) plotted on a velocity
scale are shown in Figure~\ref{f_lw}.  (The velocity is relative to
the Local Standard of Rest, in which frame the sun is moving away from
Vela at about 12.5~km~s$^{-1}$.) The top panel is an overlay of the two
O~VI lines.  The lines are centered at $\sim +18$~km~s$^{-1}$ and the
FWHM of each line is $\sim 130$~km~s$^{-1}$.  If the emission fills the
slit uniformly, then the observed line profile is the intrinsic profile
convolved with the line spread function, which, to very good
approximation, is a 106~km~s$^{-1}$ wide ``tophat'' function.  Under
this assumption the O~VI lines have intrinsic FWHMs of
$\sim 100$~km~s$^{-1}$.  The surface brightnesses of the O~VI in the
FUSE LWRS spectrum lies between the HUT P1 and P2 values
(Tables~\ref{tbl_hut}, \ref{tbl_lwrs}).  However, the ratio
I$_{1038}$/I$_{1032} \simeq 0.58$ in the FUSE LWRS spectrum, which is
lower than the value at either HUT position.  Thus, when detected over
the more spatially extended LWRS aperture, resonance scattering has a
smaller effect on the O~VI emission than when detected over the HUT
aperture at P1 or P2.

The feature on the blue wing of O~VI~$\lambda$1038 is identified as
C~II~$\lambda$1037.02, a line also detected in a FUSE spectrum of an
X-ray bright knot in the center of Vela \citep{san01}.  The transition
leading to the 1037.02\AA\ line has a lower state 63~cm$^{-1}$ above
the ground state.  The companion line, C~II~$\lambda$1036.34 is a
strong ground state transition \citep{mor91} and is not detected
because it is absorbed by the interstellar gas between us and the
SNR\@.

The bottom panel of Figure~\ref{f_lw} shows an overlay of the C~III and
the O~VI~$\lambda$1032 lines.  The C~III line is significantly broader
than the O~VI line, and it has a central reversal.  By fitting the
wings of the C~III profile with gaussians we find that the observed
peaks are centered at $\sim-60$~km~s$^{-1}$ and
$\sim+80$~km~s$^{-1}$, and that the components have FWHMs
$\sim100$~km~s$^{-1}$.  C~III is a strong resonance line, and while
the observed profile is consistent with there being two kinematic
components, we expect optical depth effects to influence the line
profile.  Some idea about the intrinsic velocity distribution of low
ionization emission is gleaned from the [S~II]~$\lambda$6716
position-velocity data presented by \citet{red00}.  In an echelle
spectrum with the slit cutting across the optical filament about
2\arcmin\ south of the FUSE LWRS position, \citet{red00} detected
bright [S~II] emission around zero-velocity.  In addition to this
central component, they found fainter red-shifted and blue-shifted
peaks (see their Figure 4).  The intrinsic velocity structure of the
C~III emission is likely to be as complex, in which case the reversal
in the observed profile is due to zero-velocity emission being
self-absorbed by both filament material and the ISM along the sightline.

\subsubsection{MDRS Spectrum}

C~III~$\lambda$977 and the O~VI doublet are detected in the FUSE MDRS
spectrum, obtained at a position several arcminutes behind the bright
optical filaments.  The line fluxes are plotted against LSR velocity in
Figure~\ref{f_md}.  The two O~VI lines track each other closely
(Figure~\ref{f_md} top panel).  Each O~VI line profile has two peaks,
centered at $\sim-80$~km~s$^{-1}$ and $\sim+80$~km~s$^{-1}$.  The width
of the gap between the two components is about 90~km~s$^{-1}$, which is
much greater than the typical velocity width $\sim40$~km~s$^{-1}$ for
absorption by interstellar O~VI \citep{jen01}.  We conclude that we are
seeing two distinct emitting components (rather than a single,
self-absorbed component) along the line of sight at this location.  The
kinematic structure of the C~III emission follows that of the O~VI
emission (Figure~\ref{f_md} bottom panel).  The C~III line has two
peaks, centered at the same velocities as the O~VI lines.  The
separation between the two C~III velocity components is larger in the
the MDRS spectrum than in the LWRS spectrum.  Because of its bow shape,
and because of the lack of complex optical morphology at the location
of the MDRS aperture (Figure~\ref{f_image}), we conclude that the two
emission components are due to shocks on the front and back sides of
Knot D\@.  The blue-shifted O~VI lines have FWHMs of about
70~km~s$^{-1}$, and are broader than their red-shifted counterparts,
which are about 50~km~s$^{-1}$ wide (FWHM).  Both C~III components have
FWHMs of about 50~km~s$^{-1}$.

The surface brightnesses, corrected for interstellar extinction, of the
blue and red shifted components of the three lines are presented in
Table~\ref{tbl_mdrs}.  The blue shifted component of C~III~$\lambda$977
is about 1.3 times as bright as the red shifted component.  In
contrast, the total O~VI flux of the blue shifted component is just
half that of the red shifted component.
I$_{\rm{O~VI}}$/I$_{\rm{C~III}}$ is about 0.9 for the red shifted
component and about 0.4 for the blue shifted component.  The ratios
between the doublet lines, I$_{1038}$/I$_{1032}$ are 0.54 for the blue
shifted component and 0.56 for the red shifted component.  
These ratios are closer to the optically thin limit compared with
the values obtained for the HUT and FUSE LWRS spectra, which is
consistent with the less edge-on viewing geometry expected at this
position.

\section{Analysis and Interpretation}

\subsection{Shock Models}

To facilitate interpretation of the ultraviolet spectra, we compare the
data with shock model calculations.  We focus on the bright filament
covered by the HUT P1 and FUSE LWRS apertures, and generate a list of
line strengths relative to I$_{\rm{O~III]}} = 100$.  The two apertures
cover different regions on the sky, and the O~VI surface brightness is
lower in the FUSE LWRS spectrum.  We use the O~VI to O~III] line ratio
measured in the HUT P1 spectrum.  The line strengths of other high
ionization lines below 1200\AA, S~VI, Ne~VI] and Ne~V], are scaled
relative to I$_{\rm{O~VI}}$ based on the FUSE LWRS measurement, and
then rescaled relative to I$_{\rm{O~III]}}$ based on the HUT P1 O~VI to
O~III] line ratio.  The FUSE LWRS surface brightnesses are used for all
other lines in the FUSE bandpass.  These data are shown in column 3
of Table~\ref{tbl_mods}.

Shock models were calculated using an updated version of the code
presented by \citet{ray79}.  The updates are discussed by
\citet{ray97}, and some atomic rates for Si~III, Si~IV and S~IV lines
have been updated based on the CHIANTI database \citep{der01}.  The
model follows the emission and cooling behind a steady shock moving
into a constant density medium.  The main inputs are the shock
velocity, the pre-shock density and the elemental abundances.  Models
are presented for a range of shock velocities.  In each case the
pre-shock hydrogen number density, $n_0$ is 1~cm$^{-3}$.  (The
ultraviolet line fluxes scale linearly with pre-shock density so model
line ratios do not depend on $n_0$.) Elemental abundances, H : He : C :
N : O : Ne : Mg : Si : S : Ar : Ca : Fe : Ni = 12.00 : 11.00 :  8.55 :
7.97 : 8.79 : 8.07 : 7.58 : 7.55 : 7.21 : 6.60 : 6.36 : 7.51 :  6.25 on
a logarithmic scale, are used in the models.  Except for the Oxygen
abundance (discussed below) these are solar
abundances based on \citet{gre89} and tabulated by \citet{fer97}.  In
all the models the pre-shock magnetic field is 1$\mu$G, the temperature
of the pre-shock gas 10,000~K\@, and the gas is fully ionized.  These
parameters do not affect the ultraviolet line strengths significantly.
For each model, the calculation is followed until the post-shock gas
reaches about 1000~K, by which point the recombination zone is
complete.

The ratio of N~III]~$\lambda$1750 to O~III]~$\lambda$1664 is relatively
insensitive to the shock velocity.  For the range of shock velocities
between about 80~km~s$^{-1}$ and 120~km~s$^{-1}$, the ratio is a
measure of the relative abundances of Nitrogen and Oxygen.  There is
some amount of uncertainty in the value for solar oxygen abundance.
\citet{fer97}, from which we have taken the other abundances, lists
[O]~=~8.87.  Recently, it has been suggested that the actual value is
lower than normally assumed: \citet{hol01} gives [O]~=~8.73, and
\citet{alp01} give [O]~=~8.69.  We ran a set of 100~km~s$^{-1}$ shock
models varying the oxygen abundance between 8.69 and 8.87 (and keeping
other abundances fixed).  The observed N~III] to O~III] line ratio is
obtained for Oxygen abundance, [O]~=~8.79, and we used this value in
the models presented in columns 4--9 of Table~\ref{tbl_mods}.

The optical depths of the strongest resonance lines observed are of
order unity in the direction normal to the shock front.  When viewed
close to edge-on (as expected at and near P1), the optical depths are
higher.  A significant fraction of the line photons emitted by the
shocked gas are scattered out of the line of sight, reducing the
observed fluxes.  While comparing spectra with shock model predictions,
it is necessary to take into consideration this decrease in fluxes due
to resonance line scattering.

In the case of the O~VI lines, the ratio between the intensities of the
two lines of the doublet is a measure of the line optical depth. If we
assume that the optical depth along the line of sight is much higher
than the transverse optical depth then, as described by \cite{lon92},
we can calculate the intensity correction factor.  For an observed
ratio I$_{1038}$/I$_{1032}\simeq0.76$, the intensity correction factor
is $\sim2.14$ (for the sum of both lines).  The carbon lines,
C~III~$\lambda$977, C~II~$\lambda$1335 and C~IV~$\lambda$1549 are all
significantly affected by resonance scattering.  The flux ratio between
C~III~$\lambda$977 and C~III~$\lambda$1176 provides an estimate of the
effects of resonance scattering on the 977\AA\ line.  The models
predict a ratio of about 80, while the observed ratio is about 20,
indicating a correction factor of about 4.  This conclusion is somewhat
compromised because of uncertainties in the model predictions of the
1176\AA\ line.  We expect that N~V~$\lambda$1240 flux is less affected
by resonance scattering than any of these other lines because nitrogen
is less abundant.  Specifically, we may assume that the correction
factor for the 1240\AA\ line (actually an unresolved doublet) is not
more than the correction factor for the O~VI doublet derived above.

\subsection{Shock Properties}

The ultraviolet spectrum provides several diagnostic line ratios that
can be used to estimate the shock velocities.  The ratios
N~IV]~$\lambda$1490 to N~III]~$\lambda$1750 and O~IV]~$\lambda$1403 to
O~III]~$\lambda$1604 are particularly useful: the lines are
intercombination lines and so are not subject to optical depth effects,
and the ratios do not depend on abundances.  At the HUT P1 position,
I$_{\rm{N~IV]}}$/I$_{\rm{N~III]}}~\sim0.32$ and
I$_{\rm{O~IV]}}$/I$_{\rm{O~III]}}~\sim0.44$.  A shock velocity
$v_s~\lesssim100$~km~s$^{-1}$ is required to reproduce these observed
values.  It is clear from Table~\ref{tbl_mods} that such a low velocity
shock cannot produce the observed flux of higher ionization lines such
as N~V~$\lambda$1240 and the O~VI doublet.  If we assume that both N~V
and O~VI are equally affected by resonance scattering then
I$_{\rm{O~VI}}$/I$_{\rm{N~V}}~\sim7.5$.  To produce this ratio requires
a shock velocity of about 170~km~s$^{-1}$.  If N~V is less affected by
resonance scattering, then the ratio is higher and so is the required
shock velocity.  The ratio between Ne~VI]~$\lambda$1006 and
Ne~V]~$\lambda$1146 is less accurate because the lines are weak, but
these lines are not affected by resonance scattering, and the observed
ratio implies shock velocities in excess of 180~km~s$^{-1}$.  Finally,
we note that the ratio of optical lines
I$_{\rm{[O~III]}}$/I$_{\rm{[S~II]}}~\sim3.1$ (Table~\ref{tbl_opt}),
which requires shock velocities of at least 80~km~s$^{-1}$.  (The
forbidden line strengths are predicted by shock models but are not
presented in Table~\ref{tbl_mods}.)

From these comparisons we find that the lower ionization lines,
including O~III], O~IV], N~III] and N~IV], come predominantly from
shocks with velocities $\lesssim100$~km~s$^{-1}$ and the higher
ionization lines such as O~VI, N~V, S~VI, Ne~V] and Ne~VI] come from
faster shocks, with velocities about 180~km~s$^{-1}$.  The latter
velocity falls within the range given by \cite{bla95} for O~VI
producing shocks, but our value for the velocity of the slower shock is
less than their estimate for the C~III producing shocks (see \S1).
Intermediate velocity shocks cannot contribute a significant fraction
of the emission since that would affect all the line ratios in the
wrong way.  For example, if there were a substantial contribution from
say a 140~km~s$^{-1}$ shock, then I$_{\rm{N~IV]}}$/I$_{\rm{N~III]}}$
would be higher than observed while I$_{\rm{O~VI}}$/I$_{\rm{N~V}}$
would be lower than observed.  The lack of correlation between the
C~III and O~VI velocity distributions in the FUSE LWRS spectrum
(Figure~\ref{f_lw}) provides corroborating evidence that the two lines
arise in different shocks.  Since the ultraviolet line ratios are
similar in both HUT spectra (\S3.2), the arguments and conclusions
presented above for P1 also hold for P2.

The observed O~III]~$\lambda$1664 surface brightness, $I_{obs} =
3.1\times10^{-15}$ erg s$^{-1}$ cm$^{-2}$ arcsec$^{-2}$
(Table~\ref{tbl_hut}).  To proceed with our analysis, we make the
simplifying assumption that a 100~km~s$^{-1}$ shock is responsible for
all the O~III] emission.  A 100~km~s$^{-1}$ shock model using a
pre-shock hydrogen number density $n_{mod} = 1$~cm$^{-3}$ predicts an
O~III] line intensity
$4.5\times10^{-6}$/$2\pi$~erg~s$^{-1}$~cm$^{-2}$~sr$^{-1}$
(Table~\ref{tbl_mods}).  This is equivalent to $I_{mod} =
1.7\times10^{-17}$~erg~s$^{-1}$~cm$^{-2}$~arcsec$^{-2}$.
The model prediction scales with the pre-shock density.  Also the model
predicts the intensity emerging perpendicular to the shock front.
Since the shock front is viewed close to edge-on, the effective area of
the shock observed is larger than the aperture area.  The
$56\arcsec~\times~10\arcsec$ HUT aperture is placed parallel to the
shock front.  If the path length through the O~III] emitting gas is
$l$(\arcsec), then the ratio of the shock area to aperture area is
$l/10$.  Thus, $I_{obs}~=~n_{100}~\times~I_{mod}~\times~l/10$, where
$n_{100}$ is the pre-shock density for the 100~km~s$^{-1}$ shock
component.  Substituting the values for observed and model surface
brightnesses, we obtain: $n_{100}\times l \simeq 1800$.

Below we find that the dynamic pressure of the shocks
observed at the FUSE MDRS position,
$\rho_{0}v_{shock}^{2}\simeq2.6\times10^{-9}$~dyne~cm$^{-2}$.  Assuming
that this is the pressure driving the shocks at P1, for $v_{shock} =
100$~km~s$^{-1}$ and [He] = 11.0, we obtain
$n_{100}\simeq11$~cm$^{-3}$.  This, in turn, implies that the path
length through the emitting gas is about 164\arcsec, which is about
0.2~pc at the distance of Vela.  The length of the optical filament in
the plane of the sky is about 430\arcsec, which is significantly higher
than the derived path length through the O~III] emitting gas.  The
difference is probably due to the curvature of the shock front into the
plane of the sky, but could also be because the cloud that the shock is
running into is approximately cylindrical with its long direction
oriented in the plane of the sky.  As we showed above, the high
ionization lines are produced by faster shocks.  For a 180~km~s$^{-1}$
shock, the isobaric condition yields a pre-shock density
$n_{100}\times(100/180)^{2}\sim3.5$~cm$^{-3}$.

At the FUSE MDRS position, the observed O~VI to C~III ratios are 0.4
for the blue-shifted component and 0.9 for the red-shifted component
(Table~\ref{tbl_mdrs}).  (Note:  no correction for resonance scattering
was made in this case.) We ran a set of shock models with velocities in
the range 140--160~km~s$^{-1}$ spaced by 5~km~s$^{-1}$.  By
interpolating from the predicted line ratios we found that the observed
values for the blue-shifted and red-shifted components are satisfied
for shock velocities of $\sim151$~km~s$^{-1}$ and
$\sim157$~km~s$^{-1}$, respectively.  We note for shock
velocities from 140 to 160~km~s$^{-1}$ that $I_{\rm{O~VI}}$ increases
rapidly while I$_{\rm{C~III}}$ stays fairly constant
(Table~\ref{tbl_mods}).

If our line of sight intersects with a shock front once, then the
following holds \citep{ray97}: $I_{0}=I_{obs}v_{obs}/v_{shock}$, where
$I_{0}$ is the intensity of the shock viewed face-on, $v_{obs}$ is the
observed radial velocity (absolute value) and $v_{shock}$ is the shock
velocity.  We apply this to the C~III emission.  The observed surface
brightnesses are $21.8\times10^{-16}$ and
$16.8\times10^{-16}$~erg~s$^{-1}$~cm$^{-2}$~arcsec$^{-2}$ for the blue-
and red-shifted components (Table~\ref{tbl_mdrs}).  The central
velocities of the components are $\sim-80$~km~s$^{-1}$ and
$\sim+80$~km~s$^{-1}$.  Using these values in the equation above yields
$I_{0}(blue) \approx I_{0}(red) \simeq
1\times10^{-15}$~erg~s$^{-1}$~cm$^{-2}$~arcsec$^{-2}$.  The face-on
C~III intensity predicted by a 150~km~s$^{-1}$ shock model for
pre-shock density 1.0~cm$^{-3}$,
$I_{mod}(\rm{C~III})\simeq2\times10^{-16}$~erg~s$^{-1}$~cm$^{-2}$~arcsec$^{-2}$.
The line intensity scales with pre-shock density, so the pre-shock
density required to produce the observed C~III brightness is
$I_{0}/I_{mod}(\rm{C~III}) \sim 5$~cm$^{-3}$.  Using the values for
shock velocity and pre-shock density derived above, we find that the
pressure in the X-ray knot driving the shock ($P =
\rho_{0}v_{shock}^{2}$) is about $2.6\times10^{-9}$~dyne~cm$^{-2}$
($P/k_{\rm{B}} \sim 1.8\times10^{7}$~cm$^{-3}$~K).

The dynamical pressure of the shock derived above is about twice the
value derived from the HUT spectrum of a face-on shock near the center
of the remnant \citep{ray97}, and about seven times the value in the
X-ray region associated with the face-on shock \citep{san01}.  It is
also about 25 times the characteristic pressure of the X-ray emitting
gas derived by \citet{kah85}, corrected for the revised distance of
250~pc to Vela.  The high pressure is consistent with Knot D
originating in a bullet near the center of the remnant in that the
bullets have to have ram pressures substantially larger than the
average to punch out through the shell.  The enhanced pressure would
not be consistent with the knot being a blister on the surface, such as
formed in the model suggested by \citet{mea88} for expanding SNR
shells.  It is interesting that \citet{jen95} derived a pressure
slightly higher than our value in the region around the line of sight
towards HD 72089, which is also near some arcuate optical filaments.
They also found several high velocity components in the absorption
spectrum of the star.  The enhanced pressure suggests that the high
velocity emission and optical filaments 
at the position observed by \citet{jen95} may be related to a bullet
similar to Knot D but traveling towards us.

We have used steady flow shock models to explain the observed emission
lines.  At the MDRS location, these lines arise in a shell within the
surface of an evolving bow shock.  \citet{har87} approximate a
Herbig-Haro bow shock by a series of plane parallel, steady flow
oblique shocks; an approximation that they find valid if the cooling
time is short compared to dynamical times.  Knot D is about
30\arcmin\ ($6.7\times10^{18}$~cm) beyond the main blast wave.
Assuming that it is moving at $\sim500$~km~s$^{-1}$ \citep{asc95}, it
has taken about 4000~years for it to get to its current position since
overtaking the blast wave.  The cooling time for a 150~km~s$^{-1}$
shock running into 5~cm$^{-3}$ material is about 300 years.  The front
of the bow shock is running into a denser part of the cloud.  This
encounter results in a slower shock that gives rise to the optical
filaments as also the low ionization ultraviolet lines.

The use of steady flow models has another limitation: shocks of
180~km~s$^{-1}$ are subject to thermal instabilities \citep{che82}.
The structure and spectrum of such a shock fluctuate in time
\citep{inn87, gae88}, but the spectrum averaged over time is not
greatly affected \citep{inn92}.  The spectrograph apertures include
emission from sufficiently complex filaments that the fluctuations are
probably smoothed out.

\section{Concluding Remarks}

We have analyzed far-ultraviolet and optical observations of Knot D in
the Vela SNR, and shown that the emission is due to shocks driven into
a medium with typical interstellar cloud densities and normal elemental
abundances.  We have also found that the dynamical pressure of the
shocks is higher than the characteristic value for the remnant.  We
have estimated that the path length through the optical filaments is
shorter than their extent in the plane of the sky implying a curved
shock front.  The velocity profiles of the lines revealed by high
resolution FUSE MDRS is consistent with emission from a bow shock.  The
over-pressure in Knot D supports the idea that it is a bow shock around
a bullet or cloud that originated near the center of the remnant and has
punched through the shell.  However, there is no evidence for enhanced
abundances.  If the driver of the bow shock is a shrapnel of ejecta
\citep{asc95}, then it remains undetected.  An alternative candidate
for the driver is an accelerated cloud that originated near the center.
The existence of such clouds was suggested by \citet{mck78} to explain
some of the high velocity emission observed in SNRs.

Knot D, in spite of its unique properties, may be one of many such
features including the ones observed in the plane of the sky around
Vela, as well as others moving radially towards or away from us.  In
any case, our study provides an estimate of the properties of Knot D
that can be usefully compared with results derived from X-ray data.
The properties of Knot D can also be compared with those of other
regions to find out the extent of their similarities.

The suggestion that Knot D is associated with RXJ0852.0-4622
\citep{red02} has been made mainly on morphological grounds.  The
connection also depends on the smaller remnant being at about the same
distance as the Vela SNR\@.  The distance and also the age of
RXJ0852.0-4622 are highly uncertain \citep{dun00} and provide very few
constraints on the properties of putative blow-out regions.  We have
found that the interstellar extinction towards the optical filaments is
typical for the Vela SNR\@.  Furthermore our ultraviolet data do not
show any peculiarities (e.g.~in the velocity distribution) that suggest
a connection between the knot and the newly discovered remnant.
Therefore, we believe that Knot D is associated with the Vela SNR\@.
It will require additional studies to prove or disprove conclusively
the association between RXJ0852.0-4622 and Knot D\@.

One major issue that remains unanswered is the gap of about 1\arcmin\
between the optical filaments and the edge of the X-ray knot
(Figure~\ref{f_image}).  This is not seen in typical shock-cloud
interactions (see references in \S3.1) where the X-ray edge and optical
emission are co-incident.  In those cases, the relationship of the two
components is well explained by a model in which the blast wave hits
the cloud and a reverse shock is propagated back into the remnant
\citep{gra95}.  In Knot D, the gap must be a consequence of the
dynamics of the bow shock.  The gap region could contain gas that is
hot but with a temperature not high enough to make X-rays.  The region
may then be a source of bright O~VI emission.  It could also be a
transient region of low pressure as predicted by models of thermal
instability \citep{inn92}.  A third possibility is that it is a region
of cool gas supported by magnetic pressure.  Each of these
possibilities has interesting implications for the structure and
evolution of astrophysical bow shocks, but the existing data do not
allow us to distinguish among them.  However, it is worth noting that
the spatial extent of the gap may be related to the cooling length of
the shocked gas.  At the distance of Vela, 1\arcmin\ is about $2\times
10^{17}$~cm, which is roughly the cooling length scale of a
240~km~s$^{-1}$ shock in a medium with density 3~cm$^{-3}$.  These
conditions may have obtained in the course of the evolution of the bow
shock.  Detailed models and direct observations are needed for further
elucidation.

\acknowledgements

We thank Paul Plucinsky for useful discussions, and for providing us
with the Chandra image presented in Figure~1.  We also thank Matt
Redman for giving us access to his [S~II] position-velocity data.
H2ools, a software package written by Steve McCandliss was used in
estimating the effects of molecular hydrogen absorption.  The referee's
comments helped improve the flow and focus of the paper.  This work has
been supported by NASA grant NAG5-10248 and NASA contract NAS5-32985,
both to the Johns Hopkins University.


\clearpage
\figcaption{(a) Three color image of the southern part of Knot D\@.
Narrowband H$\alpha$ and [O~III] are shown in red and green, respectively.
A Chandra ACIS-1 image (courtesy P. Plucinsky) is shown in blue.
The FUSE LWRS ($30\arcsec \times 30\arcsec$) aperture lies on
the optical filament while the MDRS ($4\arcsec \times 20\arcsec$)
aperture lies on an X-ray bright region $\sim3\farcm5$ away.
(b) Blow-up of the H$\alpha$ and [O~III] images showing the
bright optical filaments.  The HUT aperture positions and the FUSE LWRS
position are shown as white boxes.  North is up and East to the left in
both images.  The LWRS aperture is $30\arcsec\times30\arcsec$.
             \label{f_image} }
\figcaption{Optical spectra of the filaments at positions P1 and P2.
Spatial extractions from a longslit spectrum of regions overlapping
the HUT aperture positions were used to obtain these spectra (see
text for details).  There are no lines observed between 5100\AA\ and
6200\AA, so that region is not shown in these plots.
             \label{f_opt} }
\figcaption{HUT spectra of the filaments at position 1 (top panel) and
position 2 (bottom panel).  Note the difference in the range of the
y-axis scale between the two plots.  The aperture positions are shown
in Figure \protect{\ref{f_image}}.
             \label{f_hut} }
\figcaption{FUSE LWRS spectrum of the filament: the top, middle and
bottom panels show data from the SiC 2A, LiF 1A and LiF 2A channels,
respectively.  The flux range has been chosen to show the weaker lines,
and all data have been binned by 12 pixels ($\sim0.07$\AA) along the
wavelength axis.
             \label{f_fuse} }
\figcaption{Fluxes of strong lines observed in the FUSE LWRS spectrum,
plotted against velocity.
Top panel: overlay of the two O~VI lines.  The excess emission
seen on the blue wing of the 1037\AA\ line is identified as
C~II~$\lambda$1037 emission (see text for details).
Bottom panel: overlay of O~VI~$\lambda$1032 and C~III~$\lambda$977.
             \label{f_lw} }
\figcaption{Fluxes of lines observed in the FUSE MDRS spectrum,
plotted against velocity.  Top panel: overlay of the two O~VI lines.
Bottom panel: overlay of O~VI~$\lambda$1032 and C~III~$\lambda$977.
             \label{f_md} }

\clearpage
\begin{deluxetable}{lccccc}

\tablecaption{Optical Surface Brightnesses at HUT P1 and P2
                                   \label{tbl_opt}}
\tablewidth{0pt}
\tablehead{
  \colhead{Line ID} & \colhead{$\lambda$(\AA)} & \colhead{SB(P1)} &
  \colhead{SB(P2)} & \colhead{Red. Corr.}
}
\startdata
            H$\beta$   &  4861  &    9.1  & $<$1.0   &  1.40  \\
            {[O III]}  &  4959  &   25.9  &    9.2   &  1.38  \\
           {[Fe III]}  &  4987  &    2.4  & \nodata  &  1.38  \\
            {[O III]}  &  5007  &   80.7  &   28.7   &  1.38  \\
    {[Fe II]+Fe[III]}  &  5270  &    0.9  & \nodata  &  1.35  \\
             {[N II]}  &  5754  &    1.2  & \nodata  &  1.30  \\
                He I   &  5876  &    0.7  & \nodata  &  1.29  \\
              {[O I]}  &  6300  &    2.6  & \nodata  &  1.26  \\
              {[O I]}  &  6364  &    0.9  & \nodata  &  1.26  \\
             {[N II]}  &  6548  &    8.4  &    0.7   &  1.24  \\
           H$\alpha$   &  6563  &   27.5  &    2.7   &  1.24  \\
             {[N II]}  &  6583  &   26.5  &    2.3   &  1.24  \\
             {[S II]}  &  6716  &   19.7  &    2.5   &  1.23  \\
             {[S II]}  &  6731  &   14.6  &    1.5   &  1.23  \\
\enddata
\tablecomments{Surface brightness units:
$10^{-16}$ erg s$^{-1}$ cm$^{-2}$ arcsec$^{-2}$.
Measured fluxes were divided by the area of overlap between the longslit
and the HUT aperture positions ($2\arcsec\times10\arcsec$) and corrected
for interstellar extinction.  The correction factors are presented in
the last column of the table.}
\end{deluxetable}

\clearpage
\begin{deluxetable}{lcccc}

\tablecaption{Ultraviolet Surface Brightnesses measured by HUT
                                   \label{tbl_hut}}
\tablewidth{0pt}
\tablehead{
  \colhead{Line ID} & \colhead{$\lambda$(\AA)} & \colhead{SB(P1)} & 
  \colhead{SB(P2)} & \colhead{Red. Corr.}
}
\startdata
              S VI  &   933  &    3.6  &    3.1  &  5.38  \\
              S VI  &   944  &    2.9  &    2.5  &  5.12  \\
             C III  &   977  &   42.4  &   19.7  &  4.47  \\
             N III  &   991  &   15.6  &    7.8  &  4.25  \\
              O VI  &  1032  &   22.5  &   10.0  &  3.73  \\
              O VI  &  1038  &   17.0  &    6.1  &  3.66  \\
              S IV  &  1064  &    1.8  &    0.7  &  3.42  \\
              S IV  &  1074  &    2.6  &    1.6  &  3.34  \\
    {He II+[N II]}  &  1085  &    4.5  &    1.6  &  3.25  \\
          {[Ne V]}  &  1146  &    1.3  & \nodata  &  2.88  \\
             C III  &  1176  &    2.6  &    1.4  &  2.74  \\
               N V  &  1240  &    5.2  &    3.4  &  2.52  \\
              C II  &  1335  &    4.5  &    2.2  &  2.30  \\
               O V  &  1371  & $<$0.7  & $<$0.6  &  2.24  \\
             Si IV  &  1393  &    9.3  &    4.7  &  2.21  \\
           {O IV]}  &  1403  &   13.7  &    9.1  &  2.20  \\
    {N IV]+[N IV]}  &  1485  &    3.9  &    3.0  &  2.11  \\
              C IV  &  1549  &   26.4  &   15.2  &  2.07  \\
         {[Ne IV]}  &  1602  &    0.8  &    1.2  &  2.04  \\
             He II  &  1640  &   11.4  &    3.7  &  2.03  \\
          {O III]}  &  1664  &   31.1  &   13.8  &  2.03  \\
          {N III]}  &  1750  &   12.8  &    5.5  &  2.02  \\
\enddata
\tablecomments{Surface brightness units:
$10^{-16}$ erg s$^{-1}$ cm$^{-2}$ arcsec$^{-2}$.
Measured fluxes were divided by the HUT aperture
area ($10\arcsec\times56\arcsec$) and corrected
for interstellar extinction.  The correction factors
are presented in the last column of the table.}
\end{deluxetable}

\clearpage
\begin{deluxetable}{lcc}

\tablecaption{Surface Brightnesses in the FUSE LWRS Spectrum
                                   \label{tbl_lwrs}}
\tablewidth{0pt}
\tablehead{
  \colhead{Line ID} & \colhead{$\lambda$(\AA)} & \colhead{SB}
}
\startdata
        S VI  &   933  &    1.6  \\
        S VI  &   944  &    1.0  \\
       C III  &   977  &   40.6  \\
       N III  &   991  &   11.1  \\
       Ne VI  &  1006  &    0.8  \\
       S III  &  1015  &    0.7  \\
       S III  &  1021  &    0.9  \\
        O VI  &  1032  &   17.6  \\
        O VI  &  1038  &   10.3  \\
        S IV  &  1063  &    2.2  \\
        S IV  &  1074  &    1.9  \\
       S III  &  1077  &    0.6  \\
      Si III  &  1110  &    0.3  \\
      Si III  &  1113  &    0.4  \\
       Si IV  &  1123  &    0.3  \\
       Si IV  &  1128  &    0.6  \\
        Ne V  &  1137  &    0.2  \\
        Ne V  &  1146  &    0.5  \\
       C III  &  1176  &    1.8  \\
\enddata
\tablecomments{Surface Brightness units: 
$10^{-16}$ erg s$^{-1}$ cm$^{-2}$ arcsec$^{-2}$.
Measured fluxes were divided by the LWRS aperture area
($30\arcsec\times30\arcsec$) and corrected for interstellar
extinction.  The correction factors are presented in 
Table \protect{\ref{tbl_hut}}.}
\end{deluxetable}

\clearpage
\begin{deluxetable}{lccc}

\tablecaption{Surface Brightnesses in the FUSE MDRS Spectrum
                                   \label{tbl_mdrs}}
\tablewidth{0pt}
\tablehead{
  \colhead{Line ID} & \colhead{$\lambda$(\AA)} & \colhead{SB (blue)}
 & \colhead{SB (red)}
}
\startdata
       C III  &   977  &  21.8  &  16.8  \\
        O VI  &  1032  &   5.1  &   9.8  \\
        O VI  &  1038  &   2.8  &   5.5  \\
\enddata
\tablecomments{Surface brightness units:
$10^{-16}$ erg s$^{-1}$ cm$^{-2}$ arcsec$^{-2}$.
Measured fluxes were divided by the MDRS aperture
area ($4\arcsec\times20\arcsec$) and corrected
for interstellar extinction.  The correction factors
are presented in Table \protect{\ref{tbl_hut}}.}
\end{deluxetable}

\clearpage
\begin{deluxetable}{lcccccccccc}

\tablecaption{UV Spectra Compared with Shock Models
                                   \label{tbl_mods}}
\tablewidth{0pt}
\tablehead{
  \colhead{Line ID} & \colhead{$\lambda$(\AA)} & \colhead{I (P1)} &
 \colhead{M80} & \colhead{M100} & \colhead{M120} & \colhead{M140} &
 \colhead{M160} & \colhead{M180}
}
\startdata
    S VI &  937 &   12  &    0 &    0 &    2 &  11 &  29 &   30  \\
   C III &  977 &  131  & 1073 & 1154 & 1070 & 528 & 351 &  418  \\
   N III &  991 &   36  &   86 &   88 &   93 & 101 &  67 &   67  \\
{Ne VI]} & 1006 &    4  &    0 &    0 &    0 &   0 &   2 &   15  \\
   S III & 1015 &    5  &    5 &    4 &    5 &   4 &   3 &    3  \\
    O VI & 1034 &  127  &    0 &    0 &    0 &  12 & 447 & 2221  \\
    S IV & 1070 &   13  &   5 &   7 &   9 &  12 &   7 &    8    \\
   S III & 1077 &    2  &   2 &   2 &   2 &   2 &   1 &    1    \\
  Si III & 1112 &    2  &  14 &  10 &   4 &   3 &   3 &    3    \\
   Si IV & 1125 &    3  &   3 &   4 &   2 &   1 &   1 &    1    \\
 {Ne V]\tablenotemark{a}} & 1146 &  3  &  0 &  0 &  1 &  4 &  20 &  45  \\
   C III & 1176 &    6  &  12 &  14 &  13 &   7 &   4 &    5    \\
     N V & 1240 &   17  &   0 &   1 &  14 &  81 & 229 &  184    \\
    C II & 1335 &   14  & 263 & 243 & 210 &  96 &  80 &   89    \\
     O V & 1371 & $<$2  &   0 &   0 &   1 &   4 &  20 &   32    \\
   Si IV & 1396 &   30  & 129 & 113 &  47 &  23 &  22 &   26    \\
 {O IV]} & 1403 &   44  &   9 &  49 &  85 & 131 & 183 &  191    \\
 {N IV]\tablenotemark{b}} & 1490 & 13 &  6 &  20 &  30 &  46 &  34 &  30  \\
    C IV & 1549 &   85  &  83 & 398 & 711 & 638 & 317 &  360    \\
   He II & 1640 &   37  &   4 &  19 &  44 &  38 &  33 &   36    \\
{O III]} & 1664 &  100  & 100 & 100 & 100 & 100 & 100 &  100    \\
{N III]} & 1750 &   41  &  42 &  41 &  42 &  39 &  27 &   27    \\
  \\  
{I(O III])\tablenotemark{c}} & 1664 & \nodata & 0.33 & 0.45 & 0.61 & 1.21 & 1.53 & 1.56  \\
\enddata
\tablenotetext{a} {Sum of 1137\AA\ and 1146\AA\ lines.}
\tablenotetext{b} {Includes decays from both $^3$P$_1$ and $^3$P$_2$ states;
                   the latter is a forbidden transition.}
\tablenotetext{c} {Flux emerging from the shock front 
                   $10^{-5}$~erg~s$^{-1}$~cm$^{-2}$, 
                   emitted into $2\pi$ steradians.}
\end{deluxetable}


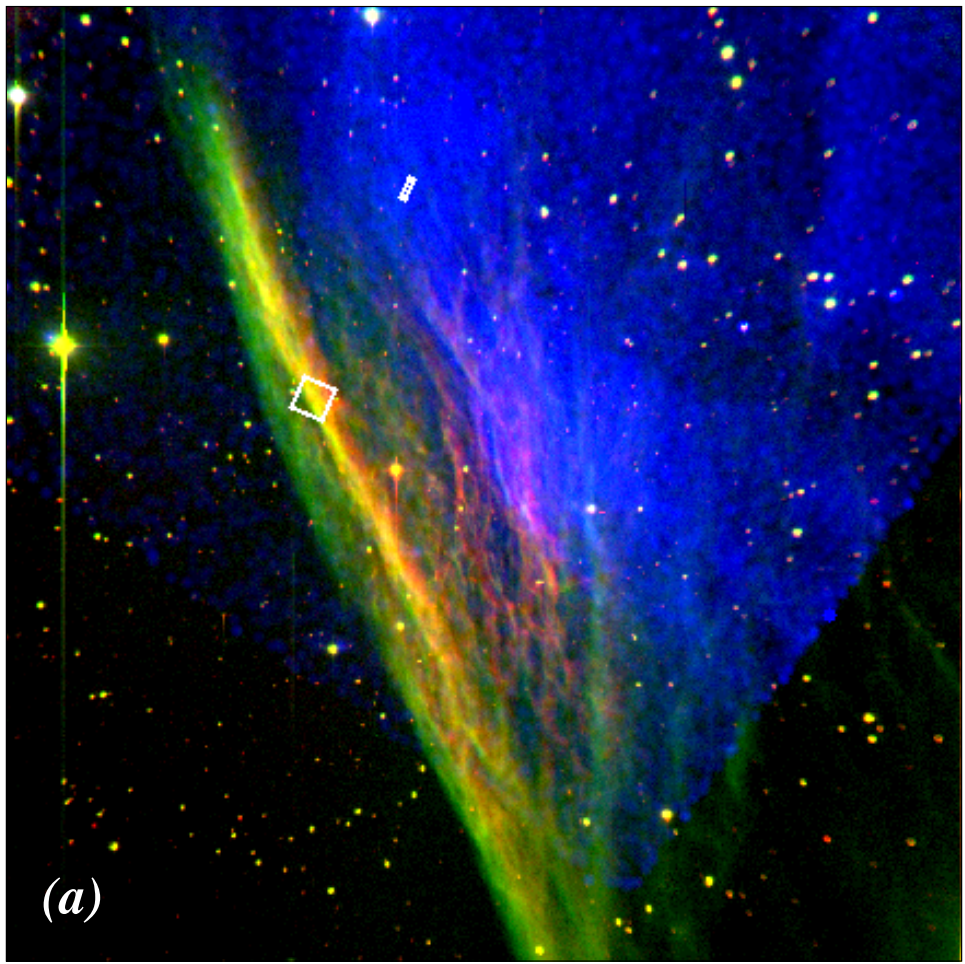
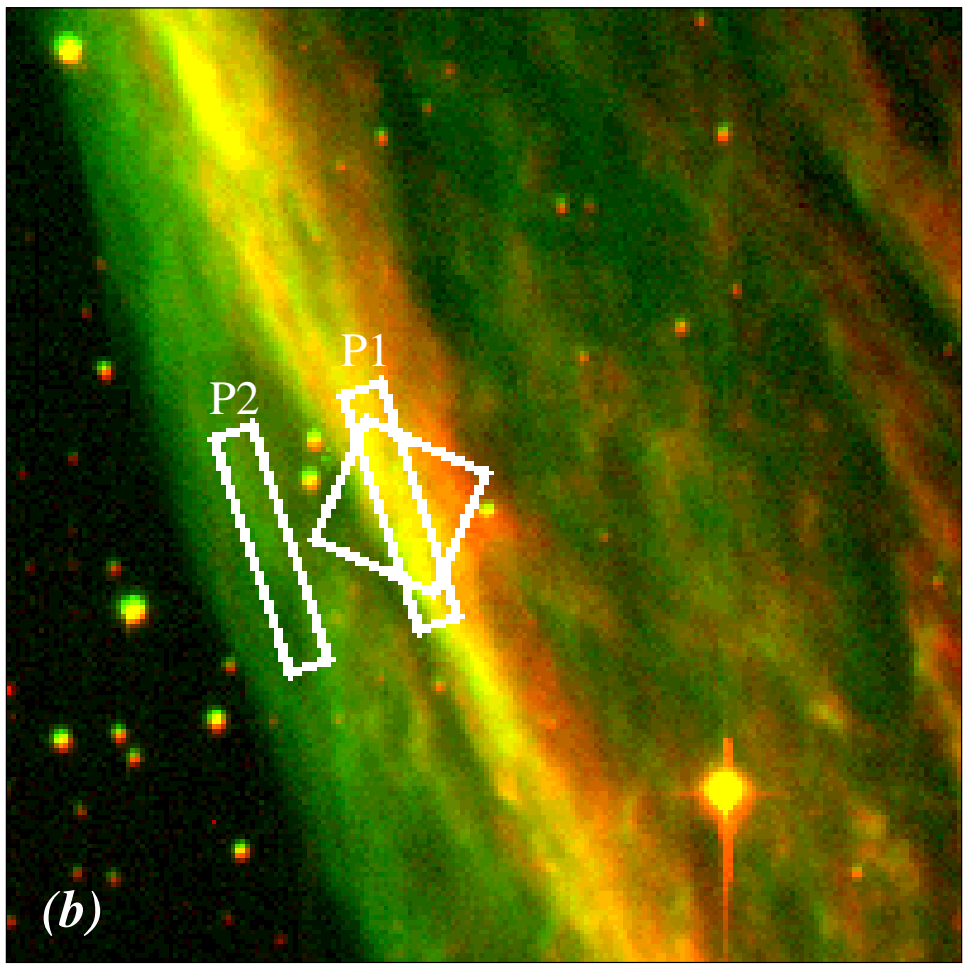
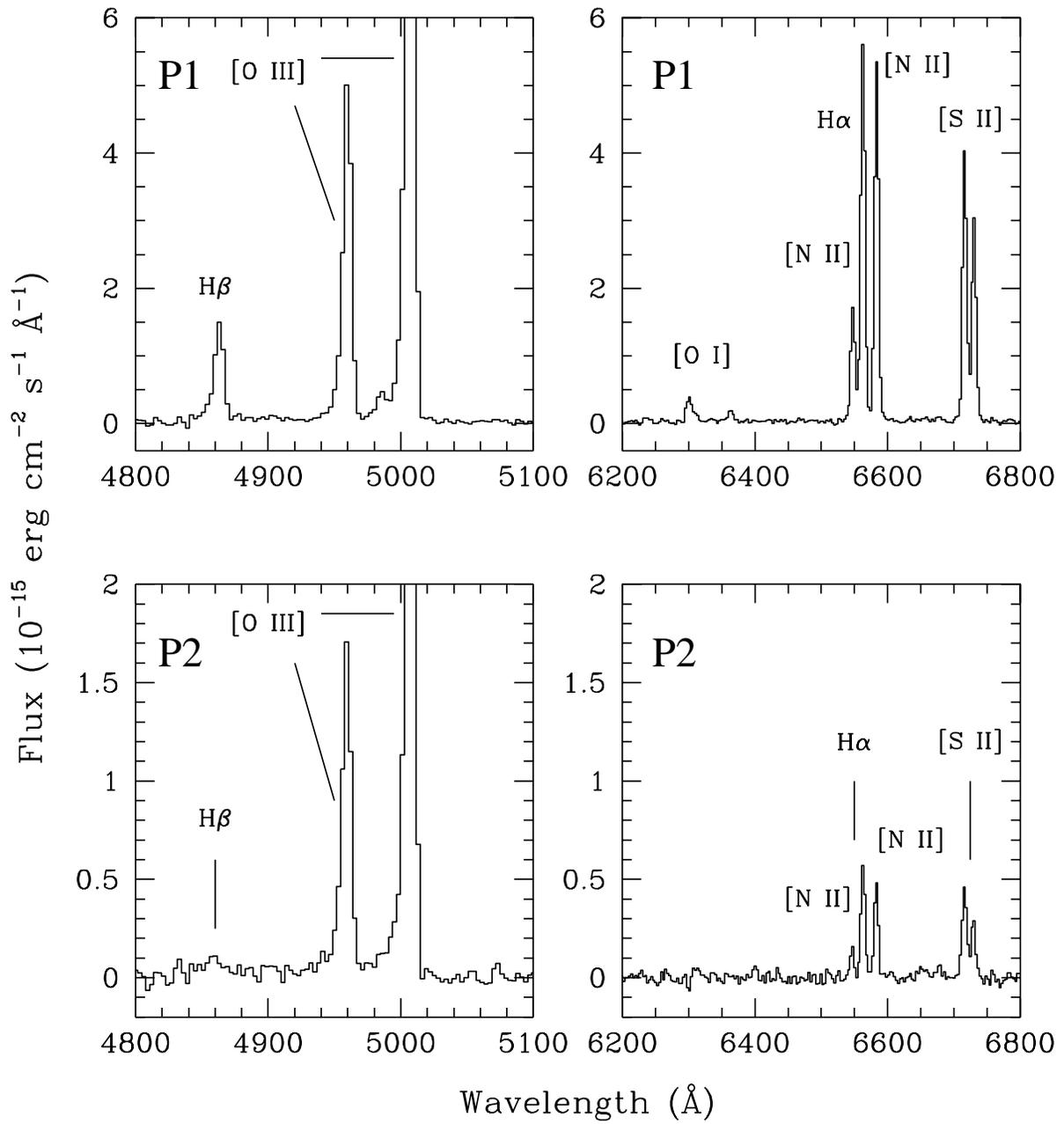
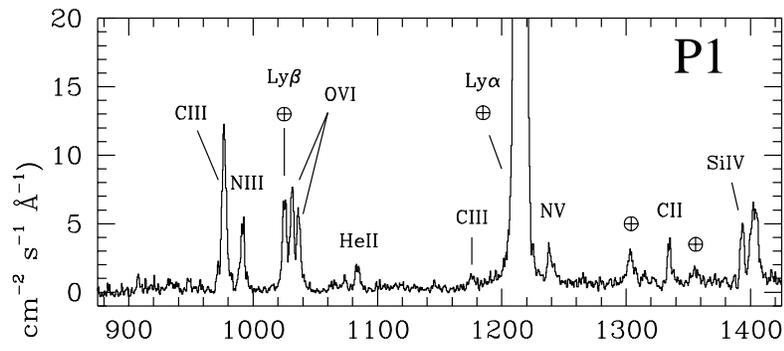
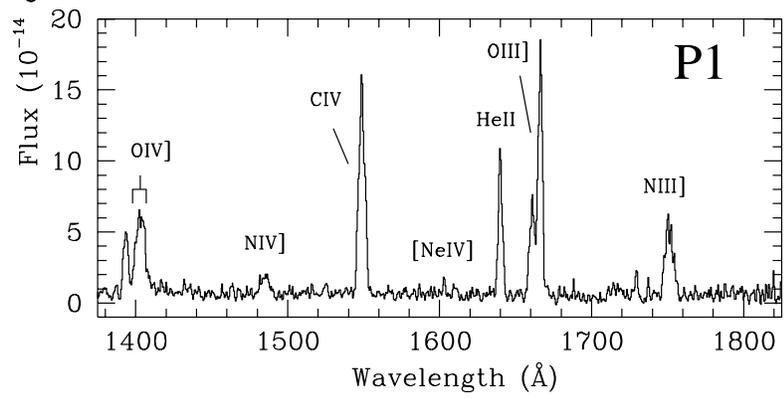
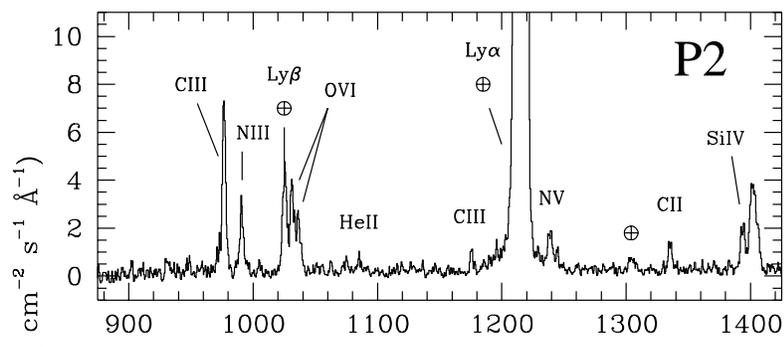
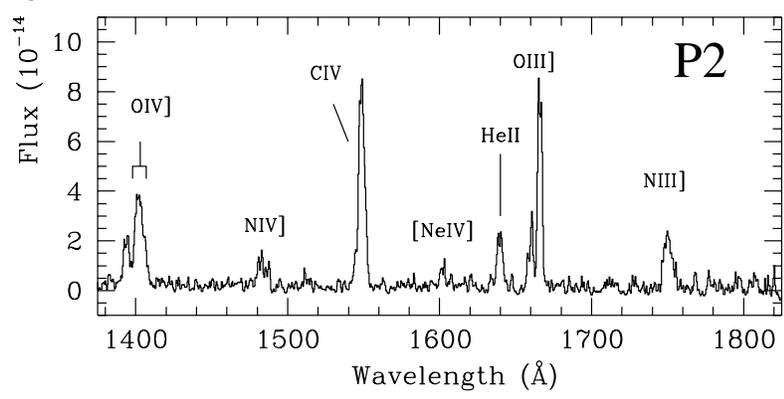
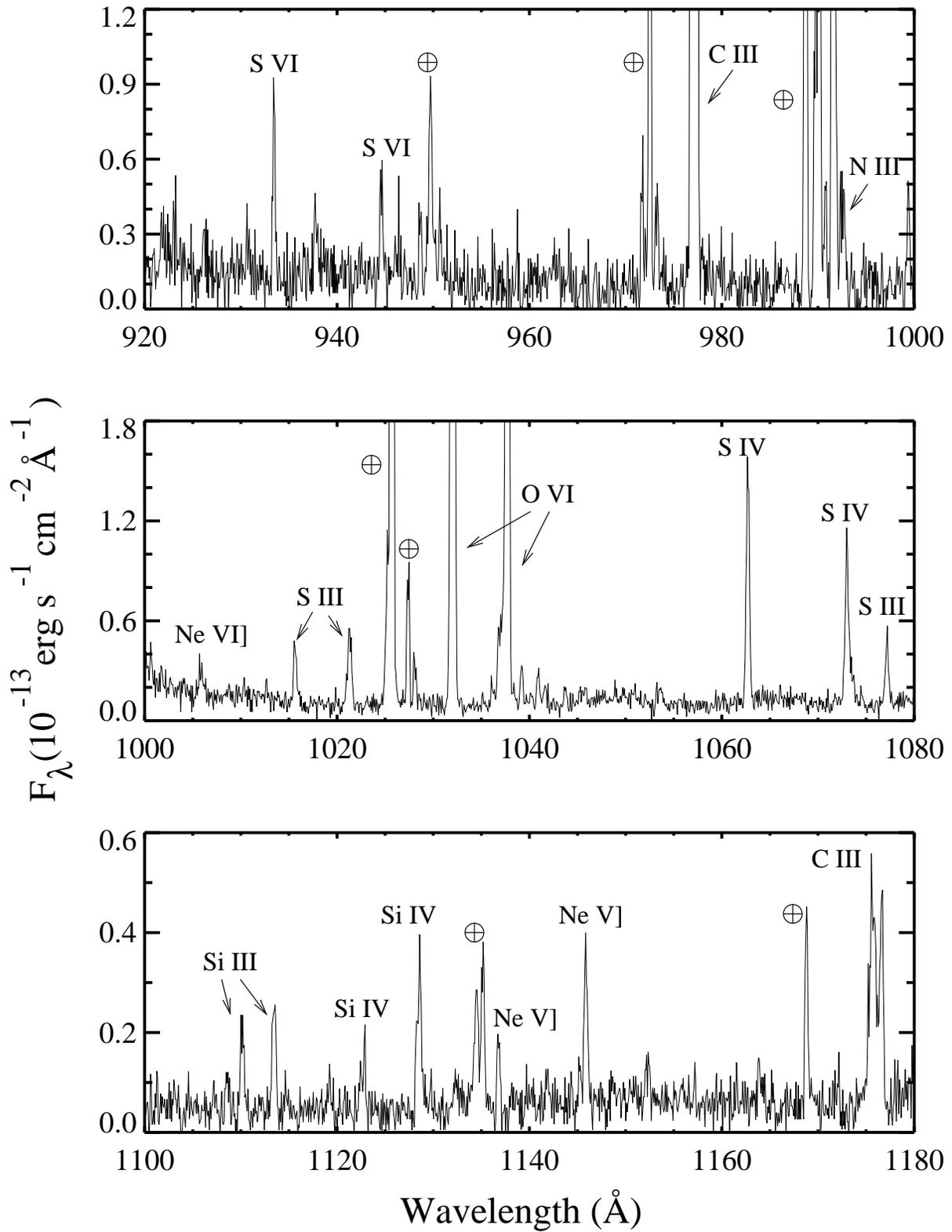
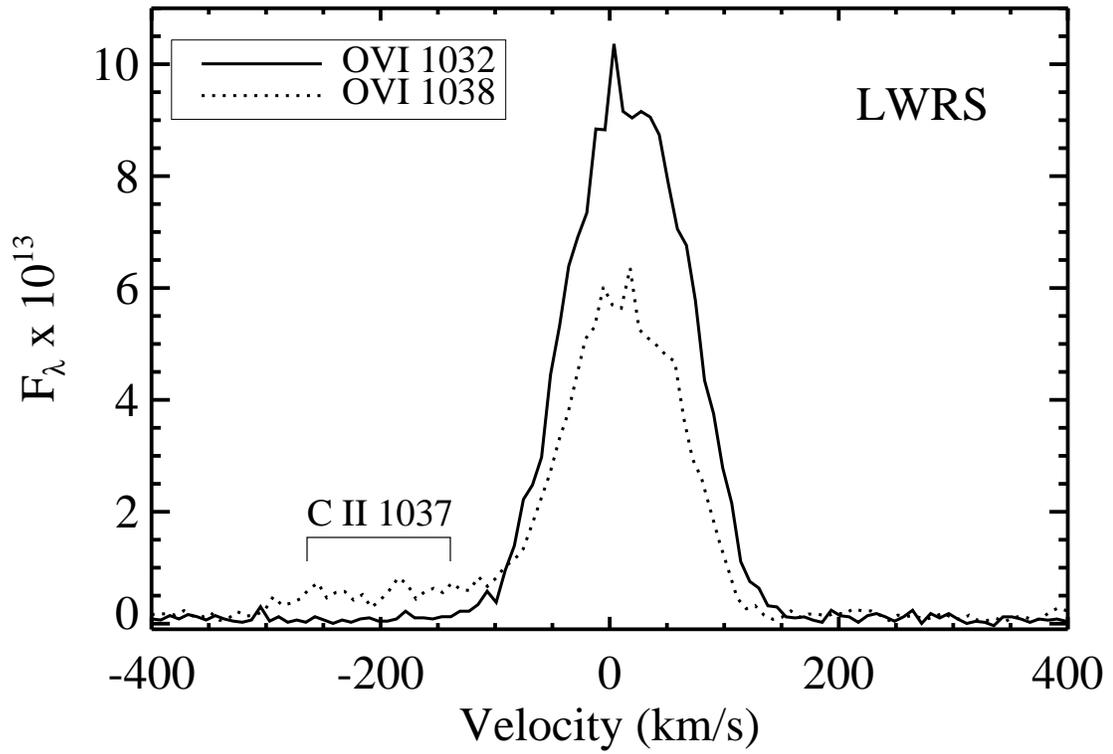
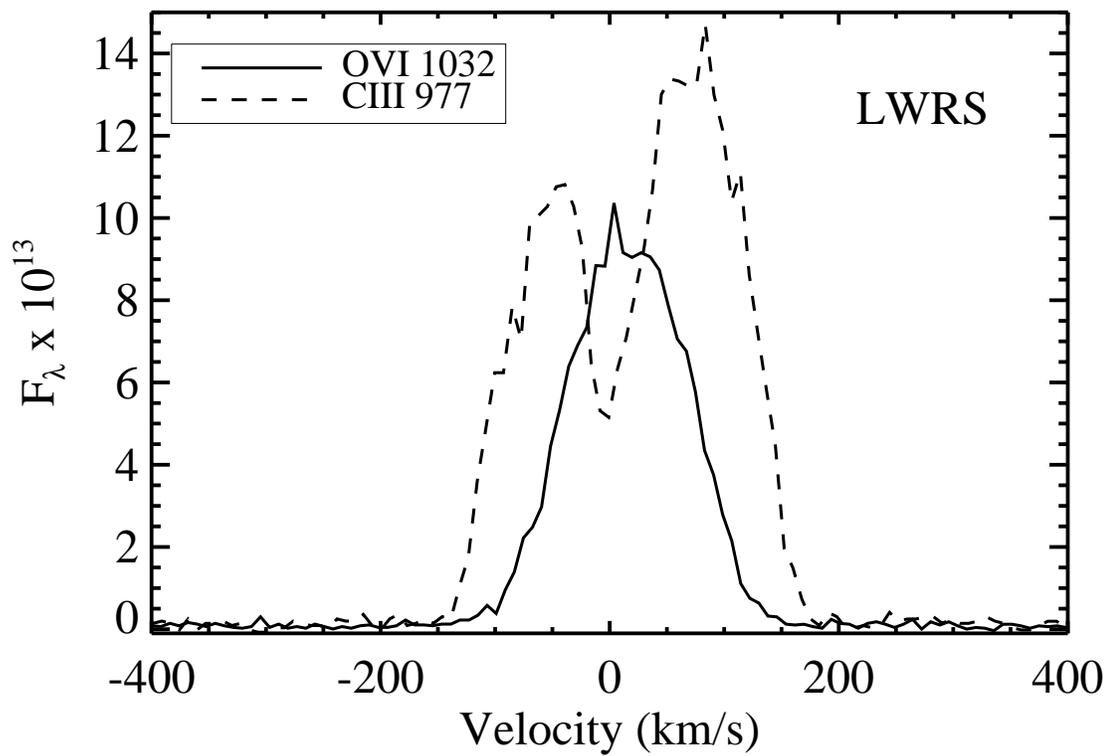
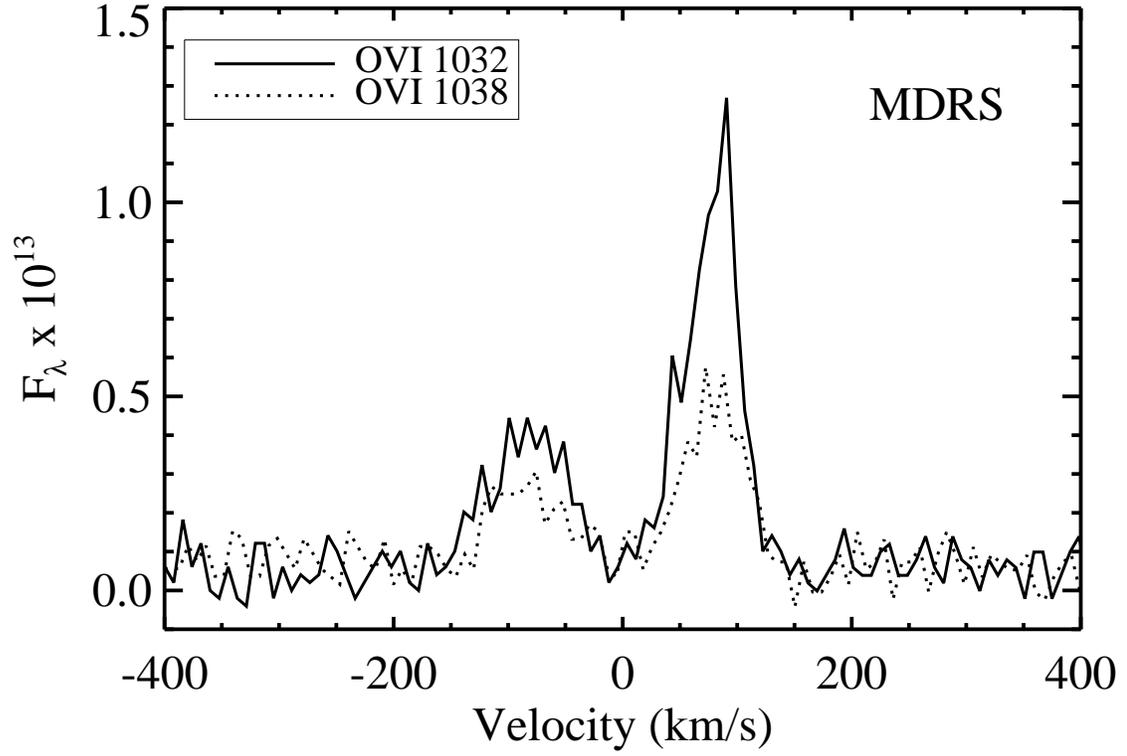
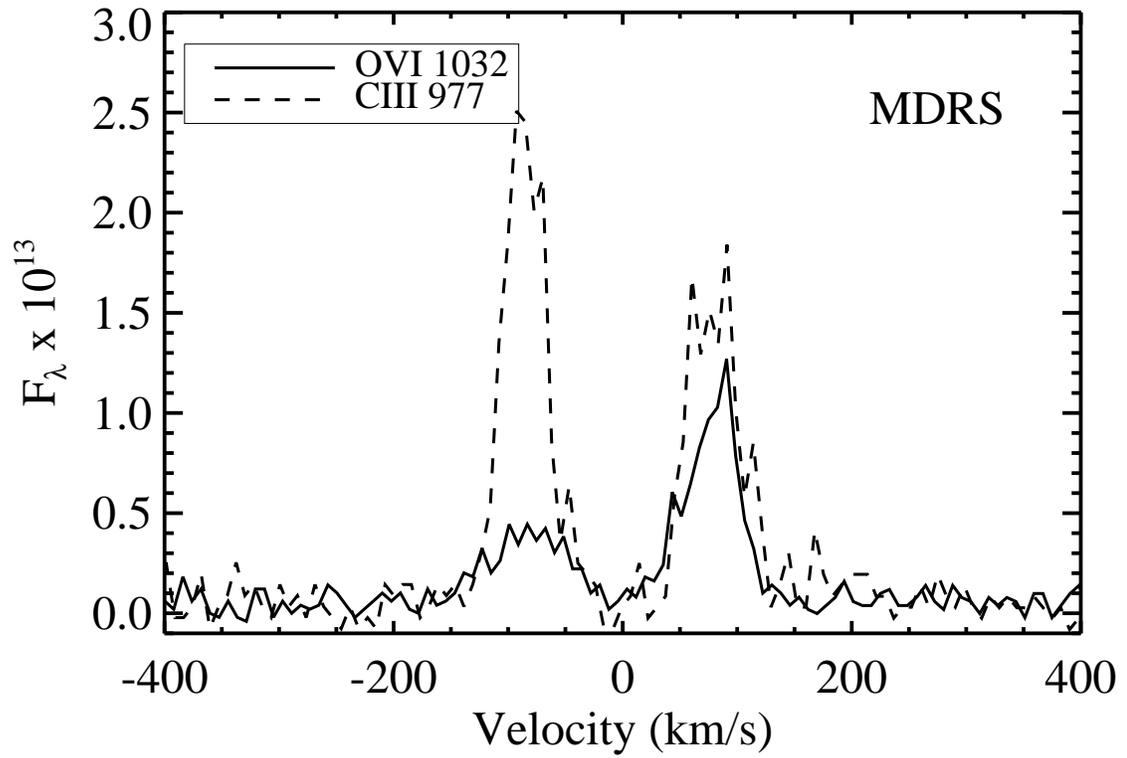
\end{document}